\documentclass[,final]{aipproc}
\layoutstyle{8x11single}
\begin{document}
\title{Noise-Induced Phase Transitions:\\Effects of the Noises' Statistics and Spectrum}
\author{Roberto R. Deza}{address={Departamento de F\'{\i}sica, Facultad de Ciencias Exactas y Naturales, Universidad Nacional de Mar del Plata, De\'an Funes 3350, 7600 Mar del Plata, Argentina.}}
\author{Horacio S. Wio}{address={Instituto de F\'{\i}sica de Cantabria (Universidad de Cantabria and CSIC), Av.\ de los Castros s/n, E-39005 Santander, Spain.}}
\author{Miguel A. Fuentes}{address={Centro Atómico Bariloche (Comisión Nacional de Energ\'{\i}a At\'omica), Av.\ Exequiel Bustillo 9500, 8400 S. C. de Bariloche, Argentina.},altaddress={Santa Fe Institute, 1399 Hyde Park Road, Santa Fe, New Mexico 87501, USA.}}
\begin{abstract}
The local, uncorrelated multiplicative noises driving a second-order, purely noise-induced, ordering phase transition (NIPT) were assumed to be Gaussian and white in the model of [Phys.\ Rev.\ Lett.\ \textbf{73}, 3395 (1994)]. The potential scientific and technological interest of this phenomenon calls for a study of the effects of the noises' statistics and spectrum. This task is facilitated if these noises are dynamically generated by means of stochastic differential equations (SDE) driven by white noises. One such case is that of Ornstein--Uhlenbeck noises which are stationary, with Gaussian pdf and a variance reduced by the self-correlation time \(\tau\), and whose effect on the NIPT phase diagram has been studied some time ago. Another such case is when the stationary pdf is a (colored) Tsallis' \(q\)--\emph{Gaussian} which, being a \emph{fat-tail} distribution for \(q>1\) and a \emph{compact-support} one for \(q<1\), allows for a controlled exploration of the effects of the departure from Gaussian statistics. As done before with stochastic resonance and other phenomena, we now exploit this tool to study---within a simple mean-field approximation and with an emphasis on the \emph{order parameter} and the ``\emph{susceptibility}''---the combined effect on NIPT of the noises' statistics and spectrum. Even for relatively small \(\tau\), it is shown that whereas fat-tail noise distributions (\(q>1\)) counteract the effect of self-correlation, compact-support ones (\(q<1\)) enhance it. Also, an interesting effect on the susceptibility is seen in the last case.
\end{abstract}
\maketitle
\section{\label{sec:1}Introduction}
As it is known, the role of temperature in equilibrium phase transitions can equally well be played by any set of spatially uncorrelated Gaussian white noises, regardless of their origin, provided that they have the same mean and variance and act \emph{additively} on the system. Under the influence of \emph{multiplicative} noises, the extended-system correlative of the phenomenon of noise-induced transitions (NIT) \cite{hole84}, namely a purely noise-induced \emph{phase} transition (NIPT), may occur. A comprehensive account of the many ways the phenomenon may take place can be found in \cite{gosa99,sago00}. However, for consistency with our previous work \cite{mdwt97,mdtw00}, we shall restrict here to the 1994 model by Van den Broeck, Parrondo and Toral (VPT) \cite{vbpt94,vptk97}, in which they proposed the following mechanism for the NIPT:
\begin{enumerate}
\item An initially unimodal pdf gets rapidly destabilized towards a multimodal one.
\item If spatial coupling is strong enough, the new states couple to form ordered domains that might subsequently coarsen and grow.
\end{enumerate}

Aiming at finding a nonequilibrium phase transition arising solely from the \emph{multiplicative} nature of the noise, and characterized (in the limit of an infinite system) by \emph{ergodicity breakdown} (only microstates compatible with the macroscopic broken symmetry should appear) and \emph{multiple} steady state probability distributions \(P^{\mathrm{st}}(\{x_i\})\), the authors in \cite{vbpt94} set up a model that is the lattice version of a scalar reaction--diffusion model submitted to \emph{multiplicative} local noises \(\eta_i(t)\). The system's state at (continuous) time \(t\) is given by \(N=L^d\) stochastic variables \(x_i(t)\) defined at the sites \(\mathbf{r}_i\) of a hypercubic lattice, and obeying a system of coupled ordinary stochastic differential equations (SDE)
\begin{equation}\label{eq:LangVPT}
\dot{x}_i=f(x_i)+g(x_i)\eta_i+\Delta_i,\quad\mbox{with}\quad f(x)=-x(1+x^2)^2\quad\mbox{and}\quad g(x)=1+x^2.
\end{equation}
Here \(\Delta_i=(D/2d)\sum_{j\in n(i)}(x_j-x_i)\) is a discretization of the Laplacian \(\nabla^2 x(\mathbf{r},t)\), \(D\) is the lattice version of the diffusion coefficient, and \(n(i)\) stands for the set of \(2d\) nearest neighbors of \(\mathbf{r}_i\). The \(\eta_i(t)\) are uncorrelated, Gaussian and white
\begin{equation}\label{eq:white}
\langle\eta_i(t)\rangle=0\quad,\quad\langle\eta_i(t)\eta_j(t')\rangle=\sigma^2\delta_{ij}\delta(t-t').
\end{equation}

As argued in \cite{vbpt94}, establishing a phase transition rigorously is a difficult task even in equilibrium, where at least the explicit form of the steady-state pdf is known. Hence, one cannot resort to the traditional techniques from equilibrium statistical physics. However, the oldest and simplest ansatz that can reproduce (albeit not always faithfully) ergodicity breakdown, namely Weiss' mean field, can be readily adapted to this nonequilibrium situation.

The method proposed by \cite{vbpt94} can be sketched as follows: By integrating the multivariable FPE over all variables except \(x_i\) and using the isotropy and translational invariance of the steady-state properties, one gets an \emph{exact} (but implicit) steady-state equation for the \emph{one-site} pdf in terms of the steady-state conditional average \(E(y)\) of \(y\in n(i)\), given the value of \(x\equiv x_i\). To determine the unknown function \(E(y)\), they introduce Weiss' mean-field (MF) approximation: to neglect fluctuations in neighboring sites, so that \(E(y)=\bar{x}\) independent of \(y\). The value of \(\bar{x}\) then follows from the self-consistency condition
\begin{equation}\label{eq:selfcons}
\bar{x}=m,\quad\mbox{with}\quad m\equiv\langle x\rangle_{\bar{x}=\mathrm{const}}
=\int_{-\infty}^{+\infty}x\,P^{\mathrm{st}}(x,\bar{x})\,dx=F(\bar{x}).
\end{equation}
When this nonlinear equation has multiple solutions, there are several corresponding steady state probabilities \(P^{\mathrm{st}}(x)\) and the MF approximation predicts a phase transition with ergodicity breakdown (usually accompanied by symmetry breakdown). If, for example, \(f\) is \emph{odd} and \(g\) \emph{even}, then any realization \(\{x_i(t)\}\) is equally probable as \(\{-x_i(t)\}\) and one should expect \(\langle x\rangle=0\). However, with the appearance of multiple solutions, this symmetry need not be fulfilled by the separate solutions, and one typically finds ``ordered'' phases with an order parameter \(m\equiv|\langle x\rangle|\neq0\).

Since \(F(\bar{x})\) is odd, \(\bar{x}=0\) is always a root of Eq.\ (\ref{eq:selfcons}). Hence, if the phase transition is second-order, its \emph{phase boundary}---which provides rich \emph{qualitative} information---lies where this root becomes unstable, i.e.\ where \(\left.(dF/d\bar{x})\right|_{\bar{x}=0}=1\). Now, the results of measurements and numerical simulations are correlations, order parameters and susceptibilities. Although the MF approximation is unable to predict the former ones, it yields predictions of the remaining two that can thus serve as a guide for numerical and (prospective) real experiments. The ``susceptibility'' we look upon is the MF correlative of the one defined in \cite{vbpt94,vptk97} for numerical simulations:
\begin{equation}\label{eq:suscept}
\chi\equiv\frac{1}{\sigma^2}\left[\int_{-\infty}^{+\infty}x^2\,P^{\mathrm{st}}(x,\bar{x})\,dx-m^2\right].
\end{equation}

\begin{figure}
\includegraphics[height=.3\textheight]{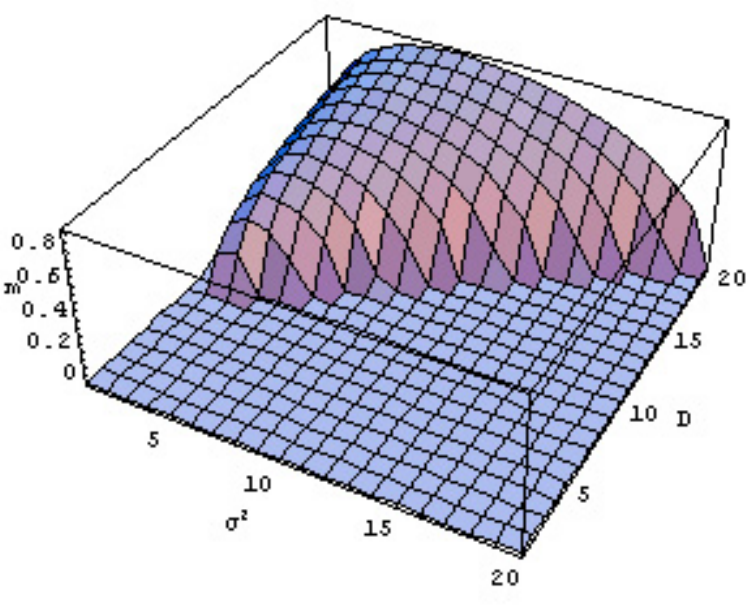}
\includegraphics[height=.3\textheight]{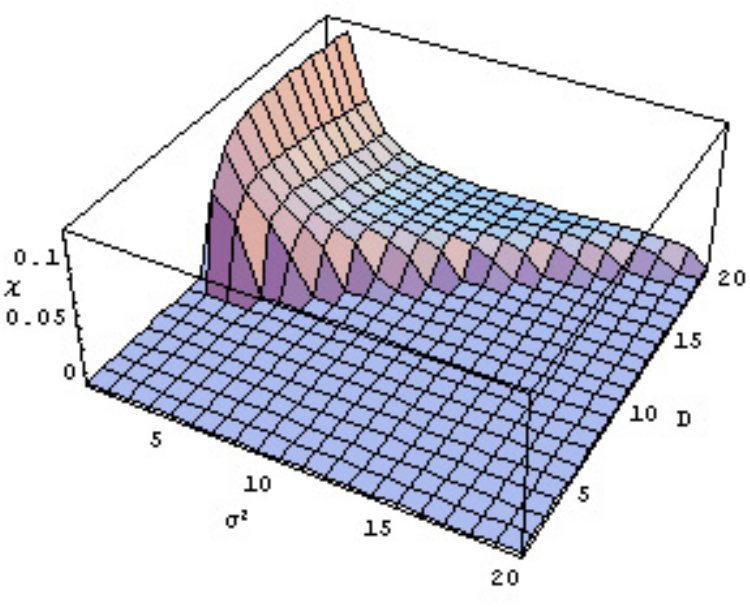}
\caption{Case of white multiplicative noise. a) order parameter and b) susceptibility as functions of coupling \(D\) and white-noise intensity \(\sigma^2\).}\label{fig:VPT}
\end{figure}

Given the exploratory character of this work, we shall perform a still simpler MF approximation consisting in replacing in Eq.\ (\ref{eq:LangVPT}) \(\Delta_i\rightarrow\bar{\Delta}_i\equiv D(\bar{x}-x_i)\), where \(\bar{x}\) is a parameter that will be determined self-consistently. Since the \(N\) SDEs get decoupled, we hereafter consider a generic one: \(\dot{x}=f(x)+g(x)\eta+D(\bar{x}-x)\).
We postpone the discussion of the numerical implementation of the MF method until we have described the process with \(q\)--Gaussian distribution. For the sake of comparison with the results obtained in that case, we illustrate the findings in \cite{vbpt94,vptk97} by plotting in Fig.\ \ref{fig:VPT} the order parameter \(m\) and the ``susceptibility'' \(\chi\) as functions of the coupling \(D\) and the white noise intensity \(\sigma^2\). That the NIPT is a \emph{combined} effect of noise and coupling is evidenced by the existence of a threshold value of \(D\). Note also that the deterministic system is \emph{disordered} however strong the coupling and that, of course, the NIPT is reentrant as a function of \(\sigma^2\).

\begin{figure}
\includegraphics[height=.3\textheight]{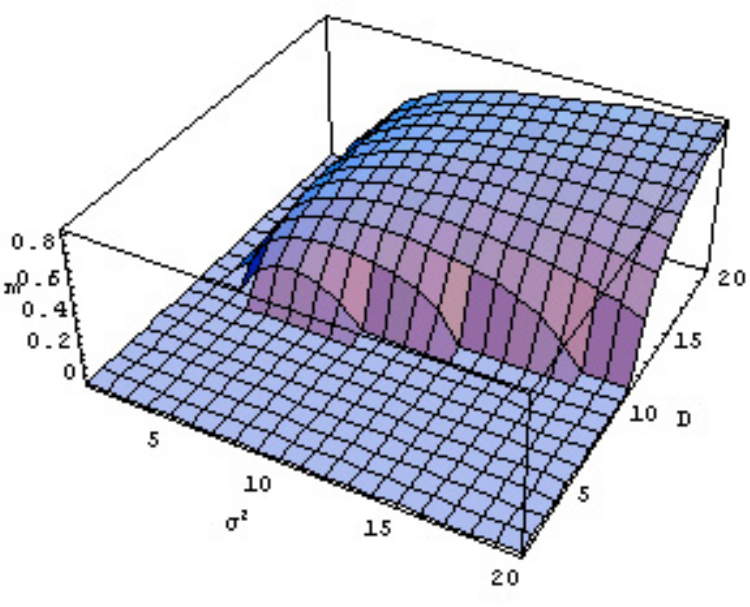}
\includegraphics[height=.3\textheight]{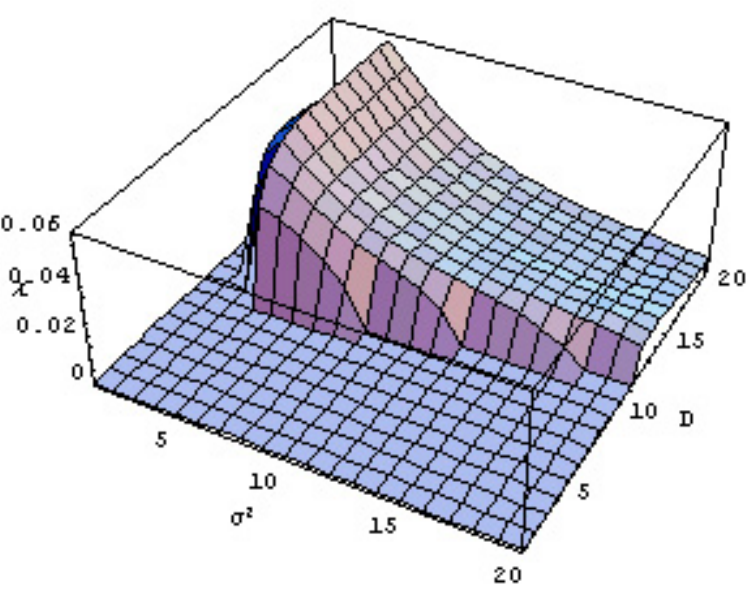}
\caption{Case of OU multiplicative noise, with \(\tau=0.1\). a) order parameter and b) susceptibility as functions of coupling \(D\) and white-noise intensity \(\sigma^2\).}\label{fig:MDWT}
\end{figure}

Early in the long list of works inspired by \cite{vbpt94,vptk97}, two of us were involved in the study of the consequences of the multiplicative noise being Ornstein--Uhlenbeck (OU) \cite{mdwt97,mdtw00}. In this case, Eq.\ (\ref{eq:white}) is replaced by \(\langle\eta_i(t)\rangle=0\) and \(\langle\eta_i(t)\eta_j(t')\rangle=\delta_{ij}(\sigma^2/2\tau)\exp(-|t-t'|/\tau)\),
where the \(\eta_i(t)\) obey
\begin{equation}\label{eq:OU}
\tau\dot{\eta}_i=-\eta_i+\xi_i,\quad\mbox{with}\quad
\langle\xi_i(t)\xi_j(t')\rangle=\sigma^2\delta_{ij}\delta(t-t').
\end{equation}
Although the colored noises \(\{\eta_i\}\) make the process
\(\{x_i\}\) non-Markovian, some approximations (interpolation
schemes) render a \emph{Markovian} (i.e.\ tractable) process, still
capturing some of the essential features. Of course, at the price of
adding a new unsystematic approximation to the MF one. However, we
dispose of a neat control parameter, namely \(\tau\), to compare
with the white-noise case. Figure \ref{fig:MDWT} is the
corresponding plot for \(\tau=0.1\). For further details, see
\cite{mdwt97,mdtw00}.
\section{\label{sec:2}Combined effect of spectrum and statistics}
One possible generalization of Eq.\ (\ref{eq:OU}) is
\begin{equation}\label{eq:13}
\tau\dot{\eta}=-\frac{d}{d\eta}V_{q}(\eta)+\xi(t),\quad\mbox{with}\quad V_{q}(\eta)=\left[\frac{\sigma^2}{2\tau(q-1)}\right]\ln\left[1+\frac{\tau(q-1)}{\sigma^2}\eta^{2}\right],
\end{equation}
proposed some time ago as model for correlated diffusion \cite{libo98}. As it occurred previously with the OU noise, this generalization provides a device to explore statistics effects by varying just one parameter (namely \(q\), at constant \(\tau\) and \(\sigma^2\); note that the proper control parameter here is \(\tau(q-1)/\sigma^2\), contrarily to the OU case in which it was \(\tau/\sigma^2\)).

\begin{figure}
\hspace{.5cm}\includegraphics[height=.25\textheight,bb= 0pt -30pt 216pt 133pt]{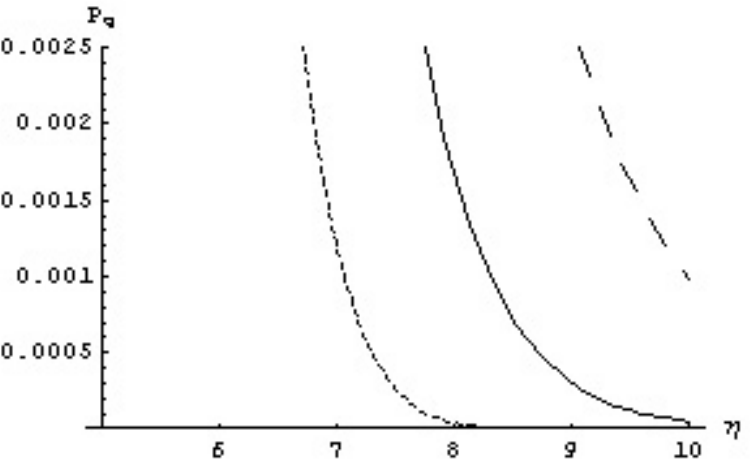}
\includegraphics[height=.3\textheight]{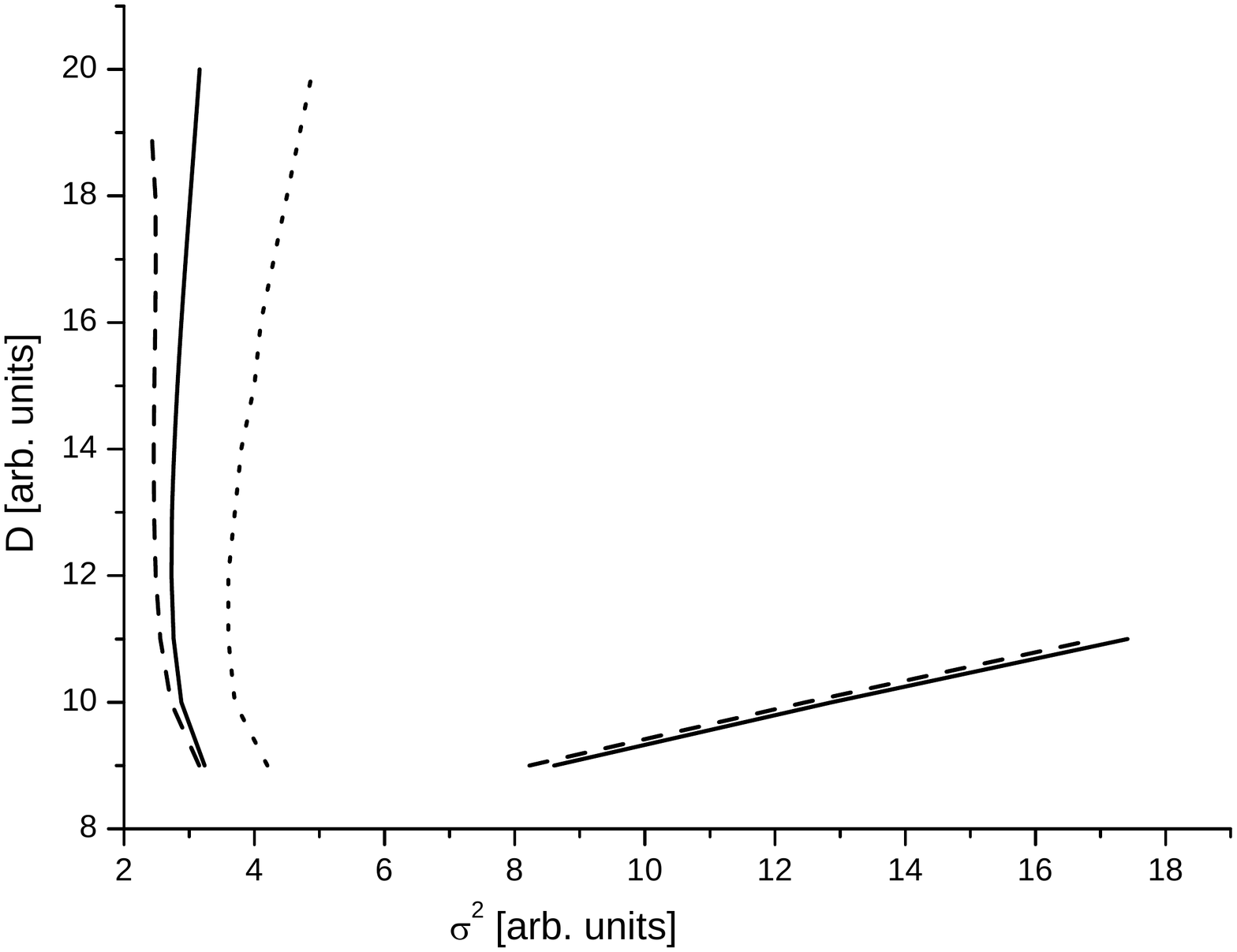}
\caption{a) Detail of the unnormalized \(P_q^{\mathrm{st}}(\eta)\) near its cutoff for \(q=0.9\), and b) phase boundaries in (\(\sigma^2\),\(D\)) plane, for \(\tau=0.1\). Dotted line: \(q=0.9\); solid line: \(q=1.0\). dashed line: \(q=1.1\).}\label{fig:q}
\end{figure}

The stationary properties of the noise \(\eta\), including the time-correlation function, have been studied in \cite{fuwt02} so here we summarize the main results. The stationary probability distribution is given by
\begin{equation}\label{eq:pdf}
P_q^{\mathrm{st}}(\eta)=\frac{1}{Z_q}\left[1+\frac{\tau}{\sigma^2}(q-1)\eta^2\right]^{\frac{1}{1-q}},
\end{equation}
where \(Z_q\) is the normalization factor. This distribution can be normalized only for \(q<3\). The first moment $\langle\eta\rangle=0$ is always equal to zero, and the second moment
\begin{equation}\label{eq:2nd}
\langle\eta^2\rangle=\frac{\sigma^2}{\tau(5-3q)}
\end{equation}
is finite only for \(q<5/3\), being larger than
\(\frac{\sigma^2}{2\tau}\) for \(q>1\). For \(q<1\) the distribution
has a cut-off, and it is only defined for \(|\eta|<\eta_c\equiv
\sqrt{\frac{\sigma^2}{\tau (1-q)}}\) (Fig.\ \ref{fig:q}a). Finally,
the correlation time \(\tau_q\) of the stationary regime of the
process \(\eta(t)\) diverges near \(q=5/3\) and it can be
approximated over the whole range of values of \(q\) as
\(\tau_q\approx 2\tau/(5-3q)\). Clearly, when \(q\to1\) we recover
the limit of \(\eta\) being a Gaussian colored noise, namely the
Ornstein--Uhlenbeck process \(\xi_{OU}(t)\), with correlations
\(\langle\xi_{OU}(t)\xi_{OU}(t')\rangle=\frac{\sigma^2}{2\tau}\exp{-|t-t'|/\tau}\)
and probability distribution
\(P^{\mathrm{st}}(\xi_{OU})=Z^{-1}\exp{-\frac{\tau}{\sigma^2}\xi_{OU}^2}\).
This process gives rise to interesting phenomena when it drives
different kinds of nonlinear systems
\cite{futw01,ckfw01,fuwt02,resw02,ftwt03,wito04,bowi04,hwio04,hwio07,bowi05,hwio05}.

The expression for \(P^{\mathrm{st}}(x,\bar{x})\) arises from a consistent Markovian approximation based on phase-space functional integration (details to be published elsewhere). The numerical implementation of the MF method compromises precision and speed: infinite integrals like those of Eqs.\ (\ref{eq:selfcons}) and (\ref{eq:suscept}) [unless \(q<1\), see below] are performed by means of a 160 pt.\ Gauss--Hermite algorithm, whereas the finite integrals in the exponent of the stationary pdf are performed by means of a 96 pt.\ Gauss--Legendre algorithm. To determine the phase boundary, the Newton--Raphson algorithm is used (unless the corresponding function is badly conditioned, in which case a succession of finer sweeps is resorted to). In the \(q<1\) case, when the integrals over \(x\) in Eqs.\ (\ref{eq:selfcons}) and (\ref{eq:suscept}) are bounded because of the bound in \(\eta\), it would be faster to solve for the \(x\)-bounds using the Newton--Raphson algorithm and then applying the 96 pt.\ Gauss--Legendre algorithm. However, the function seems not to be well conditioned for this algorithm. Hence we resort to naive integration inside a \texttt{while} loop.

The integrals performed in the aforementioned way seem to be precise enough, except for large \(D/\sigma^2\) where unphysical ordered states appear. For that reason, we have limited the exploration of the (\(\sigma^2\),\(D\)) plane to (\(\sigma^2>2\)) (except in Fig.\ \ref{fig:VPT}, where the analytical expression for \(P^{\mathrm{st}}(x,\bar{x})\) is used). Figures \ref{fig:q1,1} and \ref{fig:q0,9} exhibit the order parameter and susceptibility results respectively for \(q=1.1\) and \(q=0.9\) in the \(\tau=0.1\) case, showing that fat-tail noise distributions (\(q>1\)) \emph{counteract} the effect self-correlation (namely, they \emph{advance} the ordering boundary as \(\sigma^2\) is increased at constant \(D\)), and compact-support ones (\(q<1\)) \emph{enhance} it (they \emph{retard} the ordering boundary). Particular interest rises the effect of (\(q<1\)) multiplicative noises on the susceptibility: as seen in Fig.\ \ref{fig:q0,9}b, it shifts from being larger on the \emph{ordering} boundary to being larger on the \emph{disordering} boundary.
\begin{figure}
\includegraphics[height=.3\textheight]{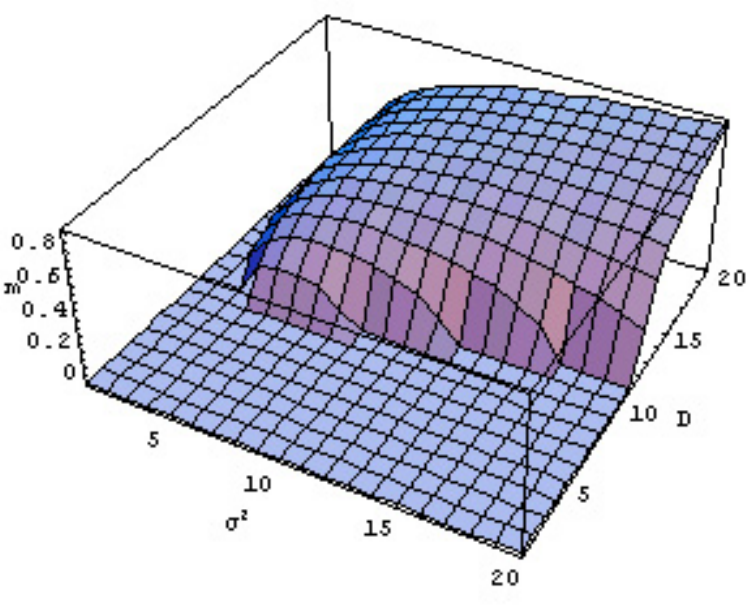}
\includegraphics[height=.3\textheight]{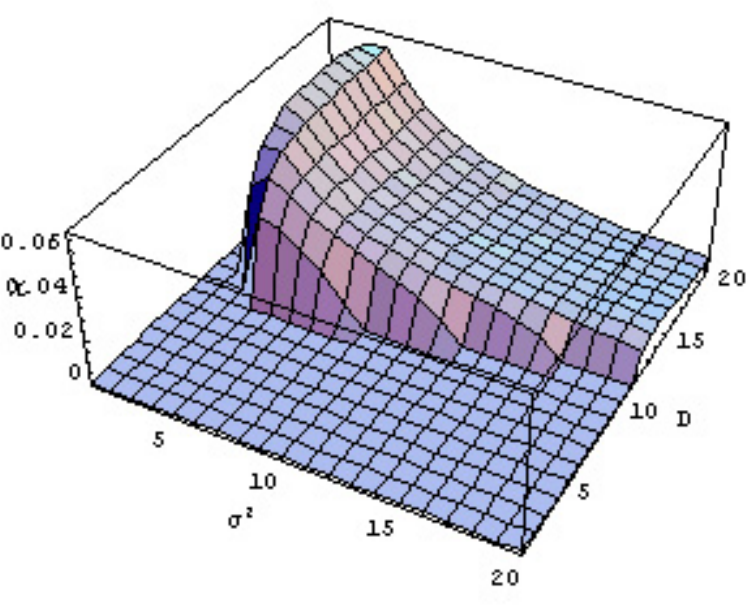}
\caption{Case of non-Gaussian multiplicative noise, with \(\tau=0.1\) and \(q=1.1\). a) order parameter and b) susceptibility as functions of coupling \(D\) and white-noise intensity \(\sigma^2\).}\label{fig:q1,1}
\end{figure}
\begin{figure}
\centering
\includegraphics[height=.3\textheight]{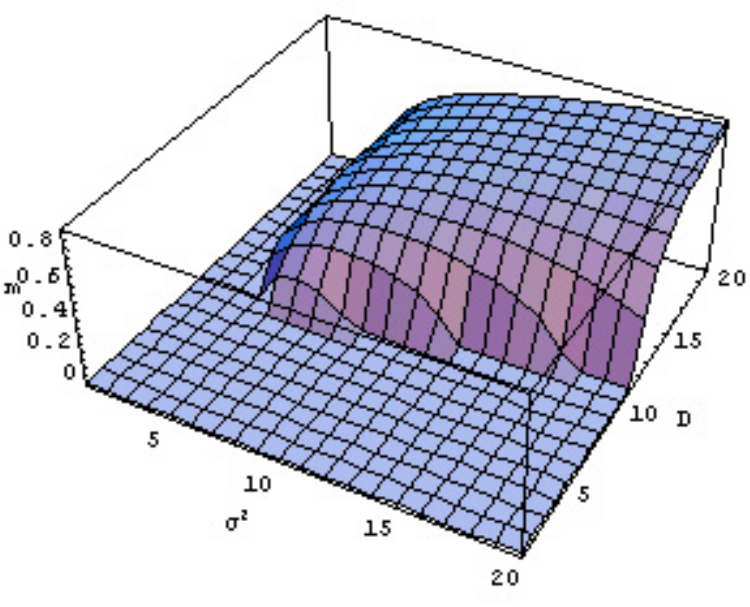}
\includegraphics[height=.3\textheight]{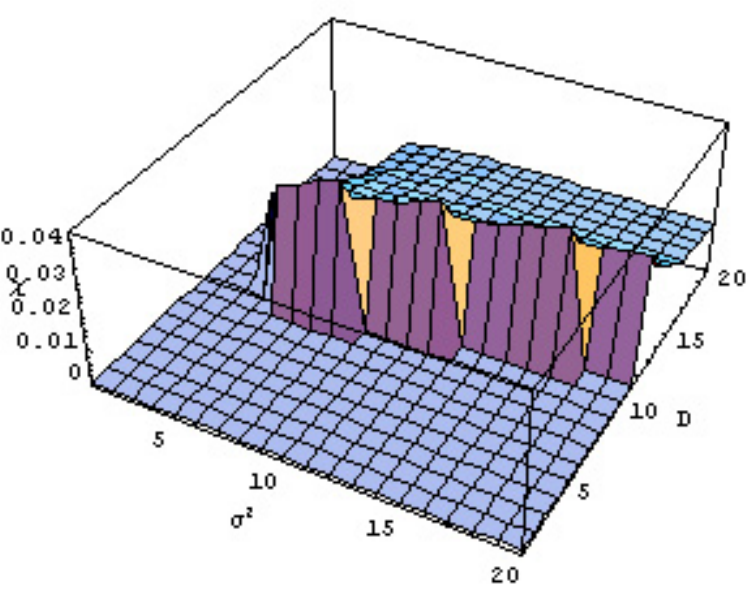}
\caption{Case of non-Gaussian multiplicative noise, with \(\tau=0.1\) and \(q=0.9\) a) order parameter and b) susceptibility as functions of coupling \(D\) and white-noise intensity \(\sigma^2\).}\label{fig:q0,9}
\end{figure}
\section{\label{sec:6}Discussion}
As warned earlier, the character of this work is exploratory, and the value of its results is to be orientative of what to expect with numerical integration of the system of SDEs, as well as with more refined MF ans\"atze and/or interpolation schemes. In order not to run into the ``forbidden'' \(\sigma^2\) region, and given the \emph{ordering} effect of \(q>1\) multiplicative noises, we have chosen to limit our exploration to \(\approx10\%\) around \(q=1\), just in order to discover trends. But in principle, our consistent Markovian approximation allows to explore the whole meaningful range of \(q\) values.
\begin{theacknowledgments}
HSW thanks the European Commission for the award of a \emph{Marie Curie Chair} at the Universidad de Cantabria, Spain. RD thanks support by ANPCyT, CONICET and UNMdP.
\end{theacknowledgments}

\end{document}